\pdfoutput=1

\documentclass[reprint,
 amsmath, amssymb,
prb,
floatfix
]{revtex4-2}

\usepackage{graphicx}
\usepackage{dcolumn}
\usepackage{bm}
\usepackage[hidelinks,
  pdftex,
  pdfauthor={Tomasz Tarkowski, Maciej Marchwiany, Nevill Gonzalez Szwacki},
  pdftitle={The structure of porous 2D boron crystals},
]{hyperref}
\usepackage{bookmark}

\usepackage[text={7.4in,9.3in},centering,
]{geometry}

\begin{document}

\title{The structure of porous 2D boron crystals}

\author{Tomasz Tarkowski}
\affiliation{Faculty of Physics, University of Warsaw, ul. Pasteura 5, PL-02093 Warsaw, Poland}

\author{Maciej Marchwiany}
\affiliation{Interdisciplinary Centre for Mathematical and Computational Modelling, University of Warsaw, ul. Pawi\'nskiego 5a, PL-02106 Warsaw, Poland}

\author{Nevill Gonzalez Szwacki}
\email{gonz@fuw.edu.pl}
\affiliation{Faculty of Physics, University of Warsaw, ul. Pasteura 5, PL-02093 Warsaw, Poland}

\begin{abstract}
In this work, we foresee the structure of a new class of borophenes with smaller 2D densities of atoms than those explored so far for 2D boron crystals. Boron atoms in the porous borophenes tend to be $5$-coordinated in contrast to commonly investigated structures with hexagonal holes for which the number of nearest neighbors of each atom varies from $3$ to $6$. High metallic character is the usual feature of borophenes, however, we have also identified a semimetallic borophene that turns into semiconductor upon unit cell expansion. A $10\%$ increase in lattice constant of this structure gives rise to an energy gap of $0.3\,\textnormal{eV}$. This extends to semiconductor industry the possible application of 2D boron crystals.
\end{abstract}

\maketitle

\section{INTRODUCTION}
The first study about 2D boron crystals is dated to 1997 \cite{Boustani1997}, however, the proposed structures were a natural extension of planar or quasi-planar boron clusters that were studied earlier by I. Boustani. At that time the 2D crystals were referred to as ``boron layers" or ``boron sheets" in which boron atoms were organized on a hexagonal lattice. Subsequent studies revealed that boron on such a lattice cannot attain planarity and undergoes structural distortions and the resulting structure is commonly referred to as buckled triangular (bt) structure. On the other hand, metal diborides became a very hot topic of study since the discovery in 2001 of high-temperature superconductivity in $\textnormal{MgB}_2$ \cite{Nagamatsu2001}. In those structures, boron atoms form graphene-like layers separated by metal atoms. The most stable structure of the 2D boron crystal is not a fully hexagonal or honeycomb structure but a ``mixture" of those two structures. Such mixed structure, which is a hexagonal atomic layer with periodically and regularly distributed hexagonal holes (atomic vacancies), is the parent structure of the proposed in 2007 boron hollow cluster (boron buckyball) \cite{GonzalezSzwacki2007}.

The formation of vacancies in the 2D hexagonal boron structure was studied in detail by using first-principles calculations at various levels. The early ones \cite{Tang2007,GonzalezSzwacki2008} were rapidly followed by studies employing systematic and sophisticated numerical methods \cite{Penev2012,Yu2012,Wu2012}. Tang \textit{et al}. \cite{Tang2007} showed that the 2D boron crystal is a polymorphic material since the distribution of hexagonal vacancies may be different for structures that exhibit similar total energies. Moreover, the authors introduced a concept of density of vacancies (hole density), $\eta$, that helps to organize the vast number of possible structures according to their value of $\eta$:
\begin{equation}
\eta=\frac{\textnormal{number of missing atoms}}{\textnormal{ number of sites in the hexagonal lattice }}.
\end{equation}

Several theoretical works, based on first-principles calculations, predicted the structure of borophene on metallic substrates \cite{Zhang2012,Liu2013,Liu2013-2,Zhang2015}. As shown by theoretical studies, various modifications of borophene can be stabilized on metal (Mg, Al, Ti, Au, and Ag) surfaces and upon adsorption on metal substrates, the hexagonal motive of the boron sheet becomes energetically competitive with respect to a triangular one. Stabilization is caused by transferring electrons from the metal to the boron sheet and the interaction between the metal and boron layer is much stronger than the interaction of the metal substrate with graphene. The substrates differ in the efficiency of interaction with the boron sheet: adsorption on Au(111) or Ag(111) substrates is accompanied by less pronounced charge transfer and interaction with the substrate than in the case of Mg(0001), Al(111), or Ti(0001) substrates \cite{Gribanova2020}.

Soon after the theoretical predictions, by using the method of laser ablation of a hot-pressed boron target, Piazza \textit{et al}. \cite{Piazza2014} provided the first experimental evidence for the viability of quasi-planar boron nanostructures with hexagonal vacancies. They showed that a $\textnormal{B}_{36}$ cluster is the smallest boron cluster to have sixfold symmetry and a central hexagonal hole. This was the first experimental evidence that single-atom layer boron sheets with hexagonal vacancies are potentially viable and the authors proposed the name of ``borophene" for an atom-thick boron crystal.

To date, there are several synthesis methods of 2D boron-based nanostructures which include: molecular beam epitaxy (MBE), liquid-phase exfoliation (LPE), and chemical vapor deposition (CVD) \cite{Tian2019}. However, MBE is by far the most suitable method for the fabrication of monolayer boron structures on metallic surfaces \cite{Ranjan2020}. Below, we describe several borophenes obtained by MBE; for a full description of the synthesis methods of 2D boron structures we refer to recent review papers \cite{Ranjan2020, Xie2020, Hou2020}.

Mannix \textit{et al}. \cite{Mannix2015} and Feng \textit{et al}. \cite{Feng2016} reported the first growth of atomically thin, crystalline 2D boron sheets on Ag(111) surfaces by MBE under ultrahigh-vacuum conditions. Mannix \textit{et al}. obtained a scanning tunneling microscope (STM) image of borophene, and ascertained that it has the same metallic and highly anisotropic characteristics as theoretically predicted \cite{Kunstmann2006} for the bt structure. Feng \textit{et al}. uncovered two allotropes of boron sheets on Ag(111) substrate with hole densities $\eta = 1/6$ and $\eta = 1/5$, which are labeled by the authors as $\beta_{12}$ and $\chi_3$, respectively. Both have a similar triangular lattice but different arrangements of the hexagonal holes as revealed by STM. Campbell \textit{et al}. \cite{Campbell2018} proved that atomically thin borophene sheets on the Ag(111) surface are bound to it by weak Van der Waals forces and determined that the thickness of a single layer borophene is $2.4\,\textnormal{\AA}$ on an unreconstructed Ag surface. Afterward, single-crystalline borophene with an area up to $100\,\mu\textnormal{m}^2$ has been fabricated on the Cu(111) substrate via the MBE method by Wu \textit{et al}. \cite{Wu2018}. The as-grown borophene was composed of novel triangular networks with a concentration of hexagonal vacancies of $\eta = 1/5$. Li \textit{et al}. \cite{Li2018} synthesized a pure honeycomb, graphene-like ($\eta = 1/3$) borophene on the Al(111) substrate by MBE. Moreover, their theoretical calculations discovered that the honeycomb borophene is stable on the Al(111) surface since there is a one-electron transfer from the Al(111) substrate to each boron atom which stabilizes the structure. The obtained honeycomb borophene has a lattice constant of about $2.9\,\textnormal{\AA}$. More recently, Kiraly \textit{et al}. \cite{Kiraly2019} selected Au(111) substrates to grow borophene via MBE and obtained islands which were confirmed by STM to be consistent with a calculated boron structure with $\eta = 1/12$ labeled as $v_{1/12}$.

The concept of regularly distributed holes in the hexagonal layer was also extended to larger holes, i.e., to borophenes with smaller 2D densities of atoms. The ``porous" boron layers were proposed to be parent structures for a whole family of boron spherical clusters \cite{GonzalezSzwacki2008}. However, until now no systematic work has been reported for porous borophenes and the present work is intended to fill this gap. 

\section{Theoretical approach}

In the present work, a full range for $\eta$ is considered ($0 \leq \eta < 1$). The search over the most stable 2D structures for a given hole density is done using a cluster-expansion-like approach in which density functional theory (DFT) energies for several 2D boron structures are used to derive an Ising-like Hamiltonian limited to nearest-neighbors (NNs) interactions, which is then used to predict the binding energies of new 2D structures. A similar theoretical model of binding energy parameterization was previously used by us in Ref.~\cite{Tarkowski2018}. In this model, the binding energy of a given 2D boron structure is decomposed into contributions of energies of the constituent atoms that have different coordination numbers. To account for 0D and 1D structures, the model was generalized by including energies corresponding to atoms with coordination number $1$ and $2$, respectively. As a result, the binding energy of a given structure can be expressed by a simple formula:
\begin{equation} \label{eq1}
E_{\textnormal{b}} \left(n_1, n_2, n_3, n_4, n_5, n_6 \right) = \frac{1}{N_{\textnormal{a}}} \sum_{i = 1}^6 n_i e_i,
\end{equation}
where $n_i$ and $e_i$ are the number of boron atoms in the unit cell with $i$ nearest neighbors and their energy, respectively, and $N_{\textnormal{a}}$ is the total number of atoms per unit cell. The individual energy contributions, $e_i$ ($i \in \{ 1, \dots , 6 \}$), are found from first-principles computations for the structure $s_i$ ($i \in \{ 1, \dots , 6 \}$) with only one type of coordination number $i$ shown in Fig.~S1. The $s_3$, $s_5$, and $s_6$ structures were previously used by us in Ref.~\cite{Tarkowski2018}. The $e_i$ ($i \in \{ 1, \dots , 6 \}$) energies are found by calculating the binding energies of the $s_i$ ($i \in \{ 1, \dots , 6 \}$) structures, respectively. Since the one-atom thick 2D boron structures are not always fully planar, the set of structures that is used to find the $e_i$ ($i \in \{ 1, \dots , 6 \}$) energies includes $5$ planar and $1$ buckled boron sheet. The values of the $e_i$ ($i \in \{ 1, \dots , 6 \}$) parameters are listed in Table~\ref{table1}.

\begin{table}
\caption{A concise description of the structures used to obtain the energy parameters employed in Eq.~\ref{eq1}. The value of the parameter $e_{i}$ corresponds to the binding energy of the $s_i$ structure. The structures are shown in Fig.~S1 and their full description is given in Tables SI and SII.}
\label{table1}
\begin{tabular}{ccc|cc}
\hline \hline
label & description             & symmetry     & parameter & \begin{tabular}[c]{@{}c@{}}value \\ (eV)\end{tabular} \\ \hline
$s_1$ & atom pair               & $\infty/mm$ & $e_1$     & $1.7803$                                                \\
$s_2$ & chain                   & $pmmm$       & $e_2$     & $5.1787$                                                \\
$s_3$ & honeycomb (hc)          & $p6/mmm$     & $e_3$     & $5.6504$                                                \\
$s_4$ & stripe (BDC)            & $pmmm$       & $e_4$     & $6.2522$                                                \\
$s_5$ & $\eta=1/7$              & $p6/m$       & $e_5$     & $6.5718$                                                \\
$s_6$ & buckled triangular (bt) & $pmmn$       & $e_6$     & $6.5116$
\\\hline \hline
\end{tabular}
\end{table}

The binding energy defined by Eq.~\ref{eq1} is used to search over the most stable structure (or structures) for a given hole density. The procedure starts by arranging boron atoms on a hexagonal lattice. Five different unit cells are considered: three with rhombus ($4\times 4$, $5\times 5$, and $6\times 6$) and two with parallelogram ($4\times 5$ and $5\times 6$) shapes. For a unit cell with $N_{\textnormal{lp}}$ ($N_{\textnormal{lp}}=16$, $25$, $30$, or $36$) lattice points and $N_{\textnormal{v}}$ missing boron atoms (vacancies), all possible arrangements of the $N_{\textnormal{a}} = N_{\textnormal{lp}} - N_{\textnormal{v}}$ atoms on a hexagonal lattice have been considered. Next, for each configuration, the $n_i$ values are determined and used to calculate the binding energies of the structures corresponding to a given hole concentration. With this method, we are able to search for the most stable configurations over the whole range of $\eta$ ($0\leq \eta < 1$) values. 

Subsequent first-principles calculations are done based on DFT and the projector augmented wave (PAW) method as implemented in the \textsc{Quantum ESPRESSO} simulation package \cite{Giannozzi2009}. For the exchange and correlation functional, we use a revised Perdew-Burke-Ernzerhof spin-polarized generalized gradient approximation (PBEsol-GGA) functional. The plane-wave basis set is converged using a $60\,\textnormal{Ry}$ energy cutoff. A $16\times 16\times 1$ \textbf{k}-point mesh and Gaussian smearing of 5~mRy is used in the Brillouin zone integration. The calculations are done using supercells ensuring a $15\,\textnormal{\AA}$ separation between adjacent layers. For each considered structure, we perform a full atomic position and lattice parameter relaxation until two consecutive relaxation steps have an energy difference less than 0.1~mRy and all residual force components are less than 1~mRy/Bohr. The images of the crystal structures are created using the VESTA visualization program \cite{Momma2011}. The space group symmetries are found by using the FINDSYM software package \cite{Stokes2005} and then converted to layer symmetry groups by using the SECTIONS software package \cite{Aroyo2006}. A $32\times 32\times 1$ \textbf{k}-point mesh is used for the electronic band structure calculation.

\section{Results and discussion}

\subsection{Predictions of the model Hamiltonian}

\begin{figure}[b]
\centering
\includegraphics[width=1.00\columnwidth]{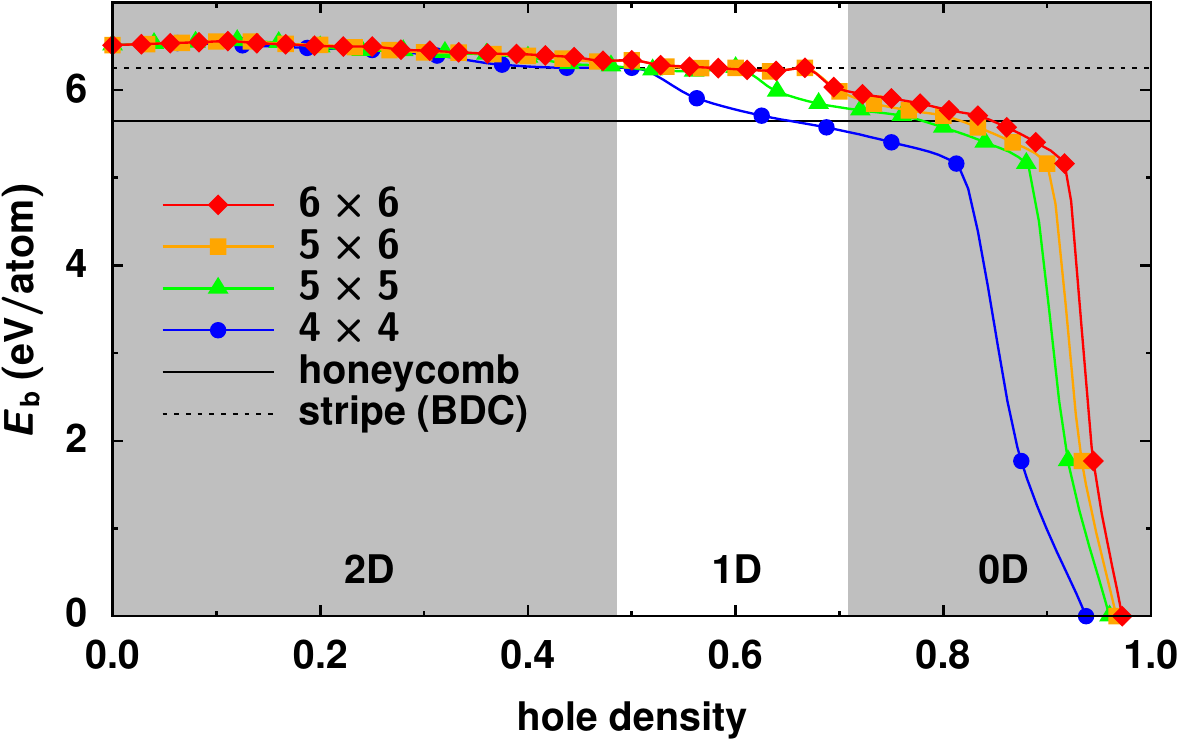}
\caption{Results of calculations based on our model. The binding energy is calculated using Eq.~\ref{eq1} and is plotted as function of hole density. Three distinct regions can be identified in the figure corresponding to structures with different dimensionality. The largest one (highlighted in gray) corresponds to 2D structures. Representatives of each region are shown in Fig.~\ref{fig2}.}
\label{fig1}
\end{figure}

\begin{figure}
\centering
\includegraphics[width=1.00\columnwidth]{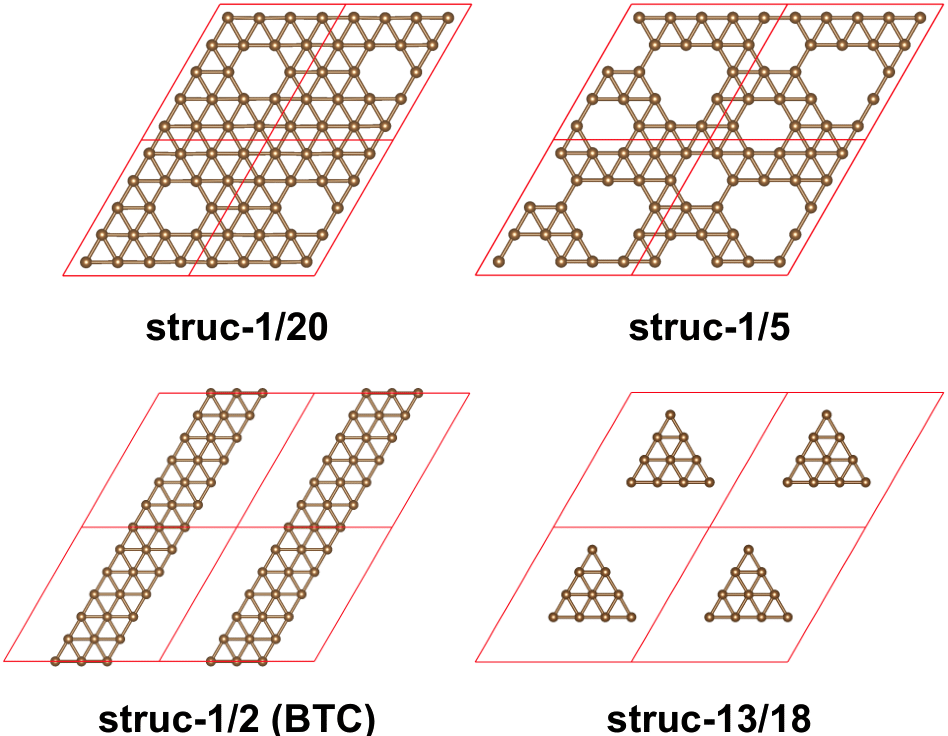}
\caption{Examples of structures found in our search using the cluster-expansion-like model. For each structure are shown $4$ unit cells each containing $4\times 5$ or $6\times 6$ lattice points. The top-left and top-right 2D structures correspond to $\eta=1/20$ and $\eta=1/5$, respectively. The bottom-left structure corresponds to $\eta=1/2$ and is a boron-triple-chain (BTC) stripe. The bottom-right structure corresponds to $\eta=13/18$ and is a cluster consisting of $10$ boron atoms.}
\label{fig2}
\end{figure}

Our model, although very simple, is able to predict the most stable structures of boron atoms arranged on a hexagonal lattice. The procedure of searching for the structures is as follows. As mentioned above, we first generate all possible structures by placing the atoms on a hexagonal lattice following all possible arrangements of $N_{\textnormal{a}}$ atoms on the available lattice points. For each of the structures, we calculate the binding energy (using Eq.~\ref{eq1}), the set of $n_i$ values, and the space group symmetry. Using those features, we identify physically nonequivalent structures. The number of resultant structures for a given hole density is usually extremely large. To filter only the most stable 2D structures, we exclude from further consideration (within DFT) 2D structures with binding energies smaller than $6.41\,\textnormal{eV}$. Our choice for the cut-off energy is not accidental, since it represents the average value of the binding energies corresponding to buckled (bt) and flat (ft) triangular structures (see Table~SII). Using our methodology, the possible candidates for the most stable structures for the whole spectrum of hole densities are known before any time-consuming DFT calculation is done. The advantage of the procedure is that it is very fast for unit cells having up to $36$ atoms ($6\times 6$).

In Fig.~\ref{fig1}, we plot the binding energies (calculated using Eq.~\ref{eq1}) for the most stable structures as a function of hole density. As a result of our search, three distinct groups of structures are found: borophenes for $0\leq \eta < 0.45$, nanoribbons for $0.45 < \eta < 0.7$, and 2D clusters for $0.7 < \eta < 1$. In Fig.~\ref{fig2}, we show examples of structures found in our search. From our analysis, we can learn that freestanding borophenes may, in principle, exist for hole densities up to about $0.45$. All those structures are more stable than the boron honeycomb and boron-double-chain (BDC) commonly known as boron stripe. The second group of structures are boron nanoribbons, and one example of that group is the boron-triple-chain (BTC) shown in Fig.~\ref{fig2} bottom-left. The last group of structures are planar boron clusters which should form if the number of missing atoms on the hexagonal lattice exceeds $70\%$. Since our model Hamiltonian was parameterized for periodic systems, we do not expect to describe well the structure of the boron cluster.

The last step of our methodology is to perform DFT calculations for all the structures that have been identified for each hole density as the most stable 2D structures in our initial search. Two separate sets of DFT calculations are done: one, in which all the structures are fully planar and the structural optimization is done preserving the planarity of the structure, and another one, in which the atoms are also allowed to move in the $z$ direction. In practice, we randomly move the atoms in the $z$ direction by not more than $\left|\Delta z\right| = 0.2\,\textnormal{\AA}$ and perform a full structural optimization of the quasi-planar structures. The buckling behavior of the 2D boron structures is discussed below. By taking the 2D structures listed in Table~SII, we may estimate that the average error (with respect to DFT values) of Eq.~\ref{eq1} in the binding energy is 45 and 54 meV/atom for buckled and flat structures, respectively.

\subsection{Structures with hexagonal holes versus porous borophenes}

\begin{figure}[b]
\centering
\includegraphics[width=1.00\columnwidth]{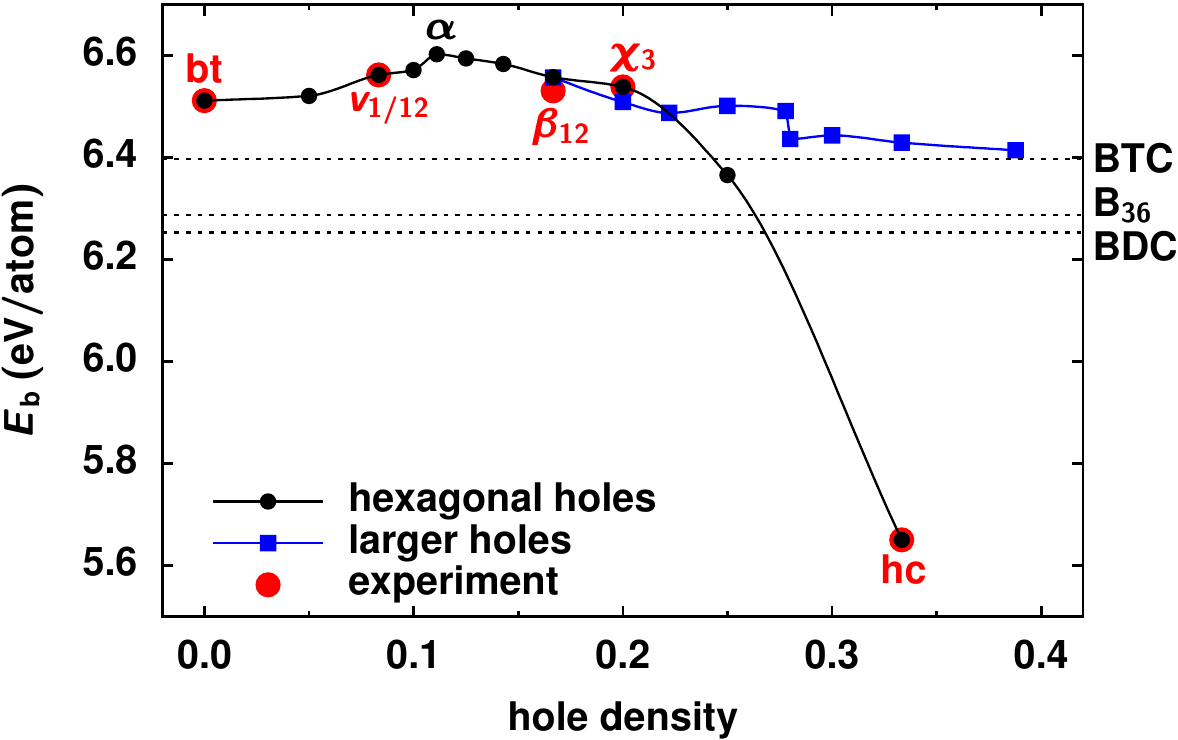}
\caption{Results of calculations based on DFT. Binding energy versus hole density for several 2D boron structures described in the text. The red solid circles correspond to structures with hexagonal vacancies that have been obtained experimentally. The structures with larger holes are shown in Fig.~\ref{fig4}.}
\label{fig3}
\end{figure}

\begin{figure}
\centering
\includegraphics[width=0.97\columnwidth]{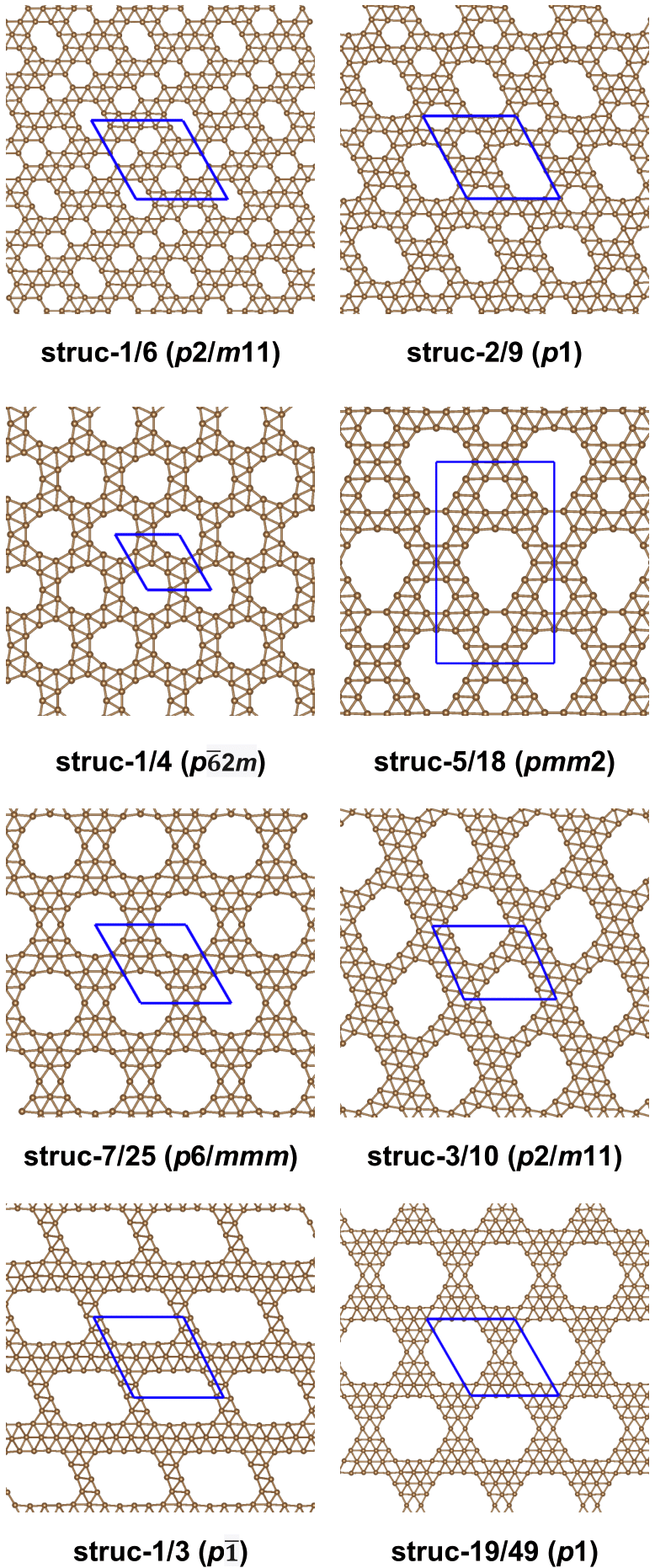}
\caption{Two-dimensional boron structures with larger holes. In parenthesis the plane symmetry groups of the structures are included. The structure struc-$1/4$ coincides with that reported in Ref.~\cite{Yu2012}, whereas structures struc-$7/25$ and struc-$19/49$ concur with those reported in Ref.~\cite{GonzalezSzwacki2008}.}
\label{fig4}
\end{figure}

The results of DFT calculations are summarized in Fig.~\ref{fig3} where we plotted the binding energy as a function of hole density for two groups of structures: one with hexagonal holes (filled circles in Fig.~\ref{fig3}) and one with larger holes (filled squares in Fig.~\ref{fig3}). It is clear from this figure that 2D structures with hexagonal holes are energetically favorable only for $0\leq \eta \leq 0.2$. For larger hole densities (up to about $0.4$) the structures exhibit larger holes (see Fig.~\ref{fig4}).

As mentioned above, most of the theoretical and experimental reports consider borophenes with isolated boron vacancies that have the shape of hexagonal holes. It is obvious, however, that the number of hexagonal holes depends upon the density of the 2D boron structure. For hole densities between $0$ and about $0.2$ the boron vacancies tend to be isolated from each other. In those structures, interwoven fragments of BDCs and BTCs separate the holes. Probably the most characteristic example of evenly distributed holes is the case of the $\alpha$-sheet \cite{Tang2007}. However, as expected for a larger density of vacancies, the boron atoms tend to aggregate, which leads to the formation of non-hexagonal larger holes. Examples of such porous sheets are shown in Fig.~\ref{fig4}. The range of hole densities for which such structures are energetically competitive spans from $0.2$ to $0.4$. The tendency for aggregation of boron atoms and the formation of triangular motives is supported by the fact that boron atoms in 2D structures tend to be $4$, $5$, or $6$ coordinated (see Fig.~\ref{fig5}). The threefold coordination is highly energetically unfavorable (see Fig.~\ref{fig3}). As a consequence, the formation of the graphene-like boron structure is not possible unless the structure is not stabilized by a metallic substrate like in a recent experimental study \cite{Li2018}. Interestingly, structures like the $\alpha$-sheet that are not only more stable but also more isotropic than $\beta_{12}$ and $\chi_3$ are not synthesized so far. The most likely reason for that is that the geometry of the substrate and the electron transfer from the metallic substrate to the 2D boron structure limits the number of boron 2D allotropes that can be synthesized \cite{Zhang2016_2}. 

\begin{figure}
\centering
\includegraphics[width=1.00\columnwidth]{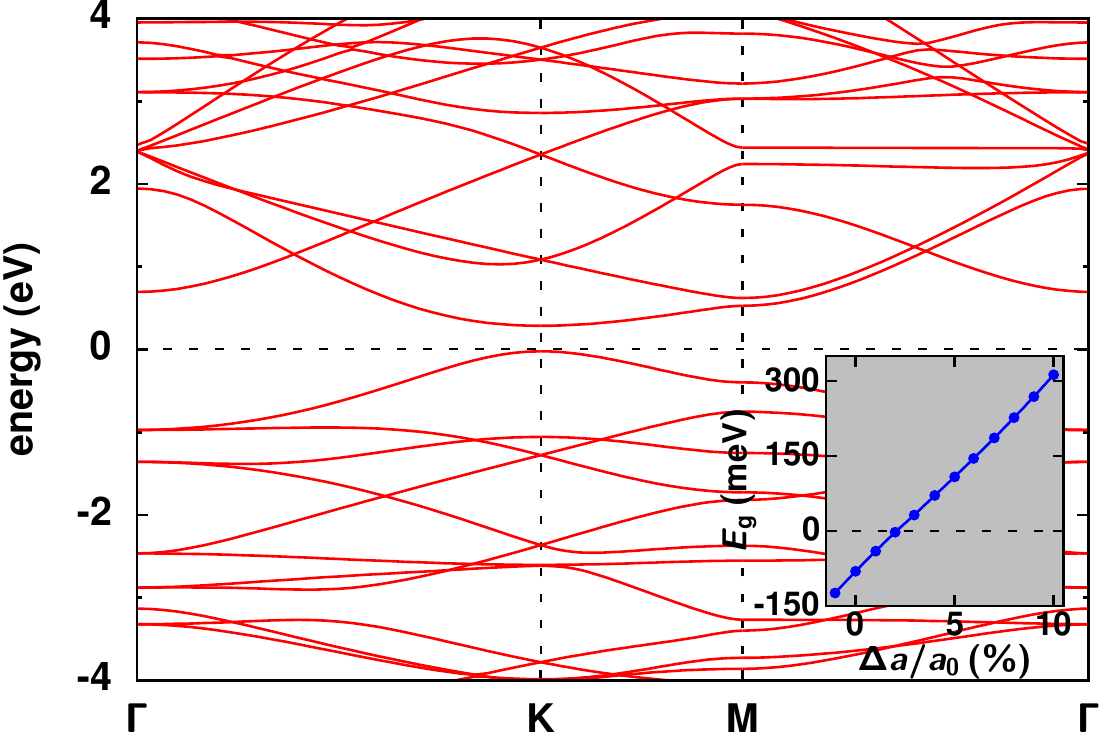}
\caption{Electronic band structure for a $10\%$ elongated lattice constant of struc-$7/25$. At the K point, VBM (CBM) is singly degenerated and corresponds to the irreducible representation $A_{1}^{\prime \prime}$ ($A_{2}^{\prime \prime}$) of the $D_{3h}$ point group. The $E_{\textnormal{g}}$ dependence on the lattice constant elongation $\Delta a / a_{0}$ is shown in the inset.}
\label{fig7}
\end{figure}

As reported in previous studies, the repeating feature of borophenes is their metallic character \cite{Ranjan2020}. However, we have also identified a structure -- struc-$7/25$ shown in Fig.~\ref{fig4} -- that is semimetallic and turns into semiconducting upon unit cell expansion. The symmetry of the structure is $p6/mmm$ and similarly as graphene has a honeycomb pattern. The overlap between the valence band and the conduction band occurs at the vicinity of the K point of the Brillouin zone. The conduction band minimum (CBM) and the valence band maximum (VBM) are located at the K point and are singly degenerated. Moreover, VBM and CBM correspond to the irreducible representations $A_{1}^{\prime \prime}$ and $A_{2}^{\prime \prime}$ of the $D_{3h}$ point group, respectively. For the relaxed structure, the energy gap  ($E_{\textnormal{g}}$) is negative with a value of $-0.081\,\textnormal{eV}$, becomes zero for an about $2\%$ lattice constant elongation and reaches a positive value of $0.313\,\textnormal{eV}$ for a $10\%$ increase of the lattice constant. The electronic band structure of struc-$7/25$ for the elongated lattice constant is shown in Fig.~\ref{fig7}, whereas the $E_{\textnormal{g}}$ dependence on the lattice constant is shown in the inset of this figure. In the case of graphene, a uniaxial or
shear strain opens a gap when it is larger than the threshold value of $20\%$ for uniaxial strain and $16\%$ for the shear strain, both values much larger than the threshold value obtained for struc-$7/25$ ($2\%$ for homogeneous expansion).

\subsection{Buckling behavior}

A common feature of many 2D boron allotropes is their tendency for buckling. This buckling behavior can be, in principle, described by using two parameters $\Delta E_{\textnormal{b}}$ and $ \Delta z $. The first of them is the binding energy difference between a structure with out-of-plane distortions and a structure confined to planarity, and the second one is the largest distance in the \textit{z} direction between the displaced boron atoms. Those two parameters may depend on many variables. Numerous variables were tested and those that in our analysis could be considered as independent are listed in Table~\ref{table2}.

\begin{table}
\centering
\caption{Set of variables considered in the machine learning data analysis.}
\label{table2}
\begin{tabular}{p{3cm}|p{5cm}}
\hline \hline
\multicolumn{1}{c|}{variable} & \multicolumn{1}{c}{description}\\
\hline
$\rho_{\textnormal{2D}}$ & 2D density of atoms (number of atoms in the unit cell, $N_{\textnormal{a}}$, divided by the area, $S$, of the unit cell)\\
\hline
$d_{\min}$ & smallest distance between atoms in the unit cell\\
\hline
$d_{\max}$ & largest distance between nearest neighboring atoms in the unit cell\\
\hline
$d$ & average distance between nearest neighboring atoms in the unit cell\\
\hline
$\sigma^{2} = \sum_{i} \frac{\left(d_{i}^{\textnormal{NN}}-d\right)^{2}}{N}$ & variance, where $d_{i}^{\textnormal{NN}}$ are the distances between nearest neighboring atoms\\
\hline
$\frac{N_{\textnormal{a}}}{N_{\textnormal{v}}}$ (and $\frac{N_{\textnormal{v}}}{N_{\textnormal{a}}}$) & the ratio between the number of atoms in the unit cell and the number of missing atoms in the hexagonal lattice\\
\hline
$\{n_{i}\}$ & number of atoms with coordination number $i$ ($i \in \{ 1, \dots , 6 \}$)\\
\hline
$n_{\textnormal{NN}}$ & average number of NNs\\
\hline
$\eta$ & hole density\\
\hline \hline
\end{tabular}
\end{table}

By using machine learning techniques based on regression, we build a simple model that describes the relationship between the buckling height and the mentioned variables:
\begin{equation}
\Delta z \left(\rho_{2\textnormal{D}}, d_{\min}, d_{\max}, d, \sigma^{2}, \frac{N_{\textnormal{a}}}{N_{\textnormal{v}}}, \frac{N_{\textnormal{v}}}{N_{\textnormal{a}}}, \{n_{i}\}, n_{\textnormal{NN}}, \eta\right).
\end{equation}

\begin{figure}
\centering
\includegraphics[width=1.00\columnwidth]{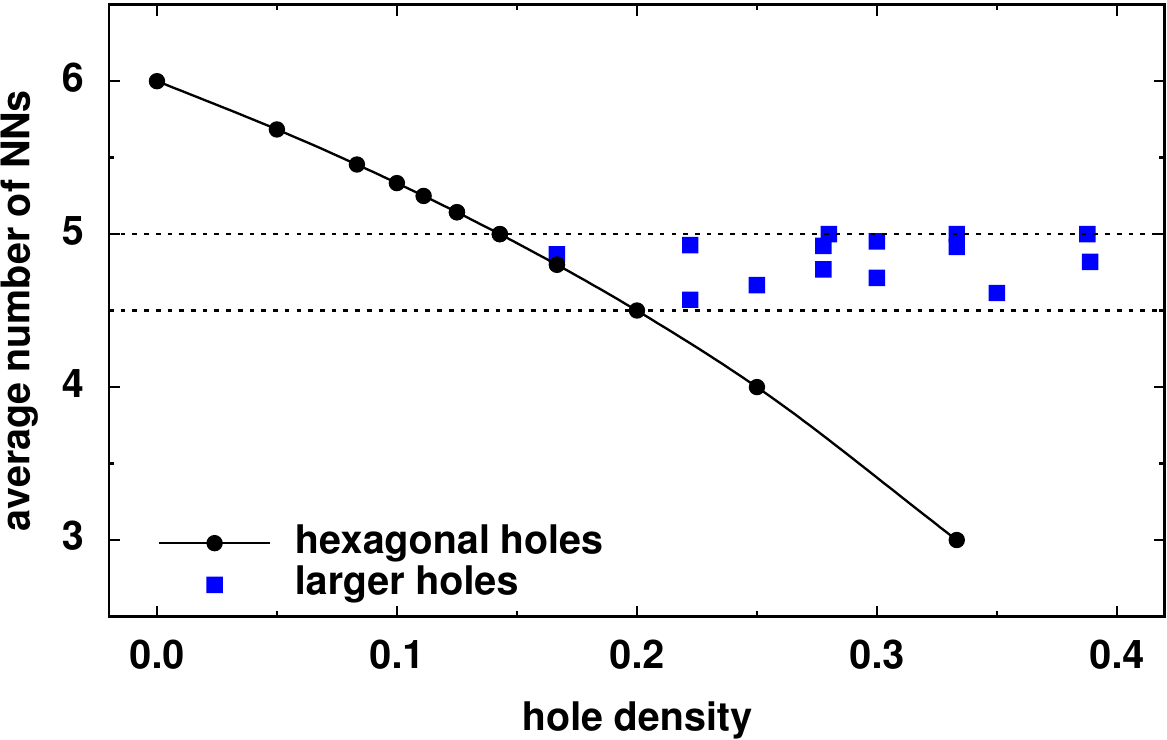}
\caption{Average number of NNs as a function of hole density. Structures with hexagonal holes exhibit much wider range of average number of NNs than structures with larger holes.}
\label{fig5}
\end{figure}

\begin{figure}[b]
\centering
\includegraphics[width=1.00\columnwidth]{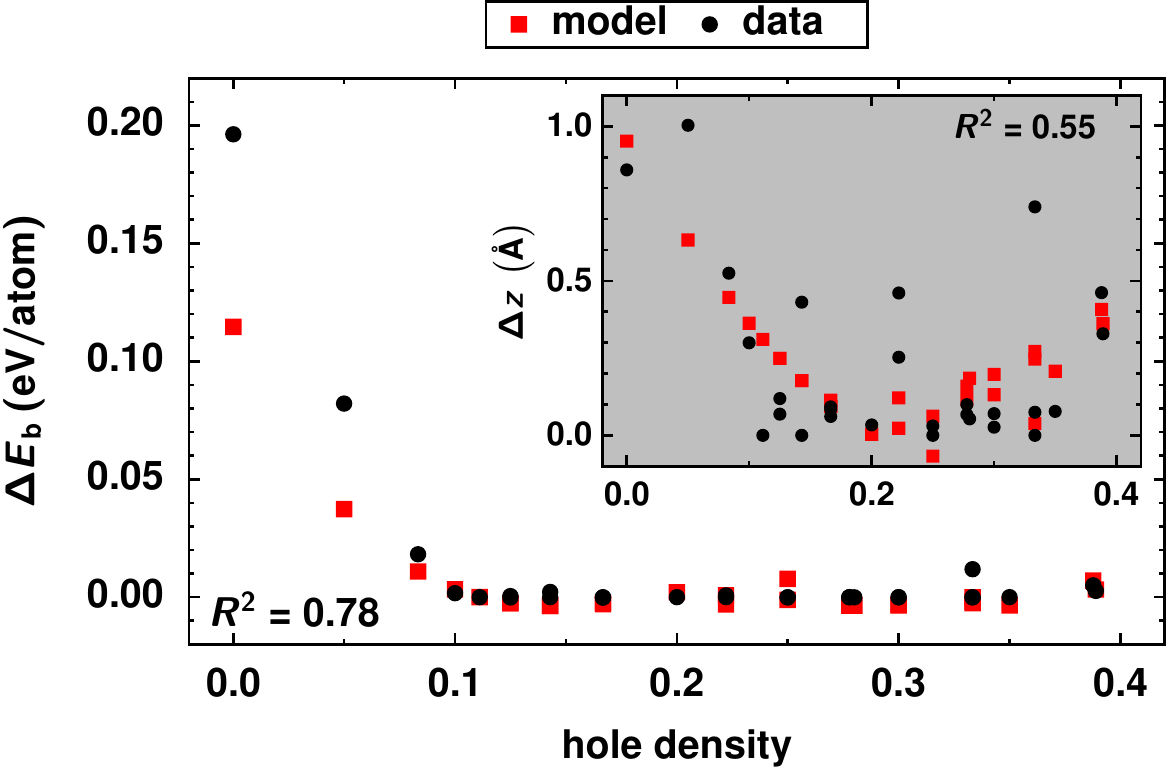}
\caption{Binding energy difference, $\Delta E_{\textnormal{b}}$, between a structure with out-of-plane distortions and a structure confined to planarity as a function of hole density $\eta$. Inset: buckling height, $\Delta z$, as a function of hole density.}
\label{fig6}
\end{figure}

The model is constructed using the data obtained for the most stable 2D boron sheets for each considered hole density complemented by several additional structures exhibiting high binding energies. The training data is, therefore, collected from $28$ structures ($14$ with hexagonal holes and the same number with larger holes, most of them listed in Table~SII). From this vast number of variables (listed in Table~\ref{table2}), we obtained that $\Delta z$ may be reasonable described by using two descriptors only, namely $n_{\textnormal{NN}}$  and $\eta$. In Fig.~\ref{fig5}, we have plotted $ n_{\textnormal{NN}}$ as a function of $\eta$. From this figure, it is clearly seen that for structures with hexagonal holes the average number of NNs varies from $3$ to $6$, whereas for structures with larger holes the variation is much narrower, namely from $4.5$ to $5$. The change of $\Delta z\left(n_{\textnormal{NN}}, \eta\right)$ with $\eta$ is shown in the inset of Fig.~\ref{fig6}. Interestingly, not only $\Delta z$ is well described but also $\Delta E_{\textnormal{b}}$. This is shown in Fig.~\ref{fig6}, where we plotted $\Delta E_{\textnormal{b}}$ as a function of $\eta$. From the figure, we may conclude that structures with $\eta < 0.1$ have $\Delta E_{\textnormal{b}} > 0$, therefore they will have a strong tendency for buckling and may be harder to stabilize at their planar forms on, for instance, metallic surfaces. On the other hand, structures with $\eta \geq 0.1$ have $\Delta E_{\textnormal{b}}\approx 0$ and, in principle, are easier to stabilize in their planar forms but as free-standing structures may undergo out-of-plane structural distortions.

\section{Summary}

Until now, only a few phases of borophene have been experimentally prepared, which is far less than the number of structures predicted by computer simulations, therefore, the synthesis of borophenes still remains a big challenge for experimentalists. We have demonstrated in this work that structures with larger holes have competitive binding energies with respect to those synthesized on metallic surfaces. Moreover, the usually observed borophenes with metallic character can be complemented with structures exhibiting semimetallic and even semiconducting properties.

\begin{acknowledgments}
We gratefully acknowledge the support of the National Science Centre under grant numbers UMO-2016/23/B/ST3/03575 and UMO-2018/31/B/ST3/03758. The use of computers at the Interdisciplinary Centre for Mathematical and Computational Modelling (ICM) at the University of Warsaw is also gratefully acknowledged.
\end{acknowledgments}

\newpage 

\bibliography{v2.bib}
\end{document}